\documentstyle[preprint,aps,epsfig]{revtex}
\begin{document}
\title{Subrecoil Raman spectroscopy of cold cesium atoms}
\author{J. Ringot, P. Szriftgiser, and J. C. Garreau}
\address{Laboratoire de Physique des Lasers, Atomes et Mol{\'e}cules and
Centre d'Etudes et de Recherches Laser et Applications,
Universit\'{e} des Sciences et Technologies de Lille\\
F-59655 Villeneuve d'Ascq Cedex, France}
\maketitle

\begin{abstract}

We describe and characterize a setup for subrecoil stimulated
Raman spectroscopy of cold cesium atoms. We study in particular the
performances of a method designed to active control and
stabilization of the magnetic fields across a cold-atom cloud
inside a small vacuum cell. The performance of the setup is monitored
by {\em copropagative-beam} stimulated Raman spectroscopy of a cold
cesium sample. The root mean-square value of the residual magnetic field is
300 $\mu G$, with a compensation 
bandwidth of 500 Hz. The shape of the observed spectra is 
theoretically interpreted and compares very well to numerically
generated spectra.
\end{abstract}


\section{Introduction}
\label{sec:Intro}
Stimulated Raman spectroscopy has become in recent years one of the most
useful and powerful tools for laser manipulation of cold atoms.
Raman stimulated transitions have been
used, for example, in subrecoil laser cooling \cite{ref:Subrecoil}; 
for the preparation of Bloch states in a stationary wave
\cite{ref:BlochOsc}; to perform 
``sideband" cooling of trapped atoms \cite{ref:Sideband}; 
and in subrecoil-precision measurement of atomic velocity distributions in
quantum chaos experiments \cite{ref:Bicolor}. This powerful technique is 
however plagued by its sensitivity to stray magnetic fields
that
should imperatively be compensated for. In the present paper, we describe
a setup used in the above mentioned quantum chaos experiment for
subrecoil-precision measurements of velocity distributions of cold
atoms, with a special emphasis in a method for reducing stray magnetic fields
to a sub-milligauss level by active compensation .

All experiments reported below were done with cesium atoms,
widely used in
the present context, but most conclusions drawn in the 
present paper can be applied to other atomic species.
Cesium recoil velocity $v_r$ (i.e. the velocity acquired by an atom emitting 
or absorbing a single photon) corresponding to the usual $D2$ line near 852 nm 
is 3.5 mm/s. The Doppler shift associated
with a stimulated Raman transition at such a velocity is 8.4 kHz.

\begin{figure}
\begin{center}
  \epsfig{figure=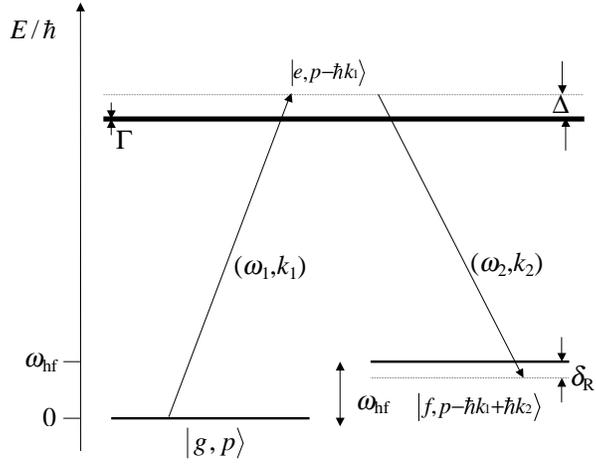,width=8cm,clip=}
\end{center}
\caption{Raman stimulated transitions.}
\label{fig:RamanTransitions}
\end{figure}

Let us briefly review some relevant features of the Raman stimulated
spectroscopy. In the present context, Raman stimulated transitions
(Fig. \ref{fig:RamanTransitions}) are
{\em two-photon} transitions connecting one hyperfine sublevel of the ground 
state to the other one by 
the absorption of a single photon in one Raman beam of frequency $\omega_1$
and wavenumber ${\bf k}_1$ and by {\em stimulated} emission of a
photon in the other beam (frequency $\omega_2$ and wavenumber ${\bf k}_2$)
{\em via} a virtual excited 
level. The Raman process is resonant if the frequency difference of the two
Raman beams and the ground-state hyperfine interval 
($\omega_{hf} \approx 2 \pi \times$ 9.2 GHz for cesium) are equal,
and the Raman detuning for zero magnetic field, low laser intensity
and zero-velocity atoms is defined as 
\begin{equation}
\delta_R = \omega_1-\omega_2 - \omega_{hf} \text{ .}
\label{eq:resraman}
\end{equation}
We suppose that the Raman beams are either parallel or
antiparallel. In what follows we shall neglect
$|k_1|-|k_2| \approx |\omega_1-\omega_2|/c$ compared to $\omega_1/c$
or $\omega_2/c$, and take $k \equiv \omega_1/c \approx \omega_2/c$.
If the Raman beams are counterpropagating (i. e. $k_1 \approx -k_2$), the
detuning seen by an atom of velocity $v$ is
$\delta_R+2k(v+v_r)$,
and the atomic velocity sensitivity of the transition will be maximum
and so is the momentum exchange between the 
Raman beams and the atom. This last property provides a way to
perform subrecoil cooling \cite{ref:Subrecoil}.
On the other hand, if the two Raman beams have the same propagation
direction (i. e. $k_1 \approx k_2$) 
the Raman detuning is almost insensitive to the atomic velocity.
As sensitivity to magnetic fields
and to light-shifts is preserved, co-propagating Raman spectroscopy is
very useful for the calibration of the apparatus.

Since the life time of the $F=4$ sublevel is very long (several thousand years 
in the absence of  collisions), the width of the Raman stimulated transitions 
will be, in principle, limited by Heisenberg principle of uncertainty.
The transitions linewidths are thus proportional to the 
inverse of the excitation duration. For instance, a Raman pulse
duration of about 20 $\mu$s, in the counter-propagating configuration,
leads to a velocity resolution of $v_r$. If the amplitude
of the pulse is adjusted to
produce a $\pi$ pulse, it will transfer a whole velocity class of 
width of $v_r$ centered at the velocity $v_0=\delta_R/2k+v_r$
from one hyperfine sublevel to the other one.
One can then reconstruct
the atomic velocity distribution by measuring the transferred
atomic population as a function of the Raman detuning $\delta_R$.
Finally, in order to prevent resonant transitions to (and 
spontaneous emission from) the excited state, the virtual
intermediate level must lie far enough 
from the closest excited level, what can be achieved by choosing an optical 
detuning $\Delta$ much larger than the linewidth of the closest excited level.
Typically $\Delta=200$ GHz in our experiment, to be compared
to the $\Gamma = 2 \pi \times 5.3$ MHz linewidth of the excited level.
The spontaneous emission rate is thus reduced by about 9 orders of
magnitude with respect to the resonance. This is true in the
limit of a narrow laser line; we have however noticed that
power diode lasers present a large spectral background about 40 dB
below the peak intensity that limits the reduction of the spontaneous
emission rate.

The optical part of our Raman setup has been described in detail elsewhere
\cite{ref:DiodeModulee}. Briefly, the Raman frequencies are generated by
direct current modulation at 4.6 GHz of a diode laser. The current
modulation generates two optical sidebands separated by 9.2 GHz that
are used to perform injection-lock of two independent power diode
lasers. We obtain in this way two 150 mW
laser beams with the required frequency separation and with a
beat-note width below 1 Hz. The Raman beams then pass
through three acousto-optical
modulators acting as optical switches that control the interaction time and 
allow shifting from the co- to the counter-propagating
beam configuration. Finally, the resulting beams are transported by
monomode, polarization maintaining, optical fibers to the region of interaction
with the atomic cloud.

\section{Active compensation of the magnetic field}
\label{sec:ActiveCompB}
\begin{figure}
\begin{center}
  \epsfig{figure=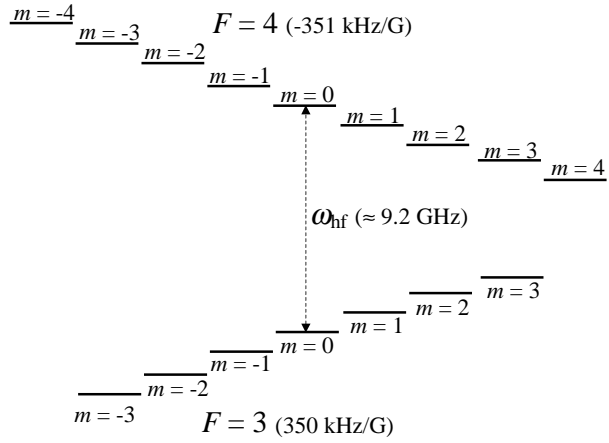,width=8cm,clip=}
\end{center}
\caption{Hyperfine structure of the cesium atom ground state.}
\label{fig:CesiumLevels}
\end{figure}

The ground state of cesium splits into two hyperfine sublevels $F=3$ and
$F=4$ separated by 9.2 GHz (the cesium-clock frequency). The
linear Zeeman split for these states is of the form $Z_F B m_F$,
where $B$ is the magnetic field, $m_F$ the azimuthal quantum number and
$Z_F$ the Zeeman coefficient (around 350 kHz/G in modulus)
(Fig. \ref{fig:CesiumLevels}). 
An atom in the $F=4$ hyperfine sublevel undergoing a 6 mG magnetic field
(2 \% of the Earth's mean magnetic field) will experience a 
Zeeman broadening equal to the recoil velocity Doppler-shift.
Thus, in order to perform atomic velocity manipulation with subrecoil 
resolution, stray magnetic fields should
be reduced to the mG level or below. Usually, this is done by shielding the
experimental zone with a double ``mu-metal" sheet providing a
high degree of isolation against magnetic fields. However, the
shield reduces the optical access to the experiment  
and, as mu-metal is difficult to tool, it also tends to forbid further
evolution of the setup. Furthermore, the magnetic gradient of
a magneto-optical trap magnetizes the shield on the scale 
of a day can lead, leading to a slow magnetic field drift.
It is thus interesting to perform an acceptable field reduction
without any ``physical" shield. In general the local magnetic field across
the experiment is a 
sum of several components: DC earth field; DC stray fields, e. g. from
an ion-pump permanent magnet; AC components from the line (50 Hz or 60 Hz)
and its harmonics; etc.
This means that the magnetic field configuration is greatly 
dependent on the particular setup one wants to protect,
and that a DC magnetic field compensation alone is insufficient.

To overcome most of those problems, the method we describe here
is to perform a real-time 
compensation of the magnetic field in the three directions with, for
each direction, DC 
and AC compensations. In order to do so, we use the fact that the sum of the
magnetic fields at the eight corners of a cube 
gives eight times the magnetic field at the center of this cube.
This result is independent of the gradient of the field, and the first
non-vanishing correction is of second order in the magnetic field
\cite{note:geometry}

\begin{figure}
\begin{center}
  \epsfig{figure=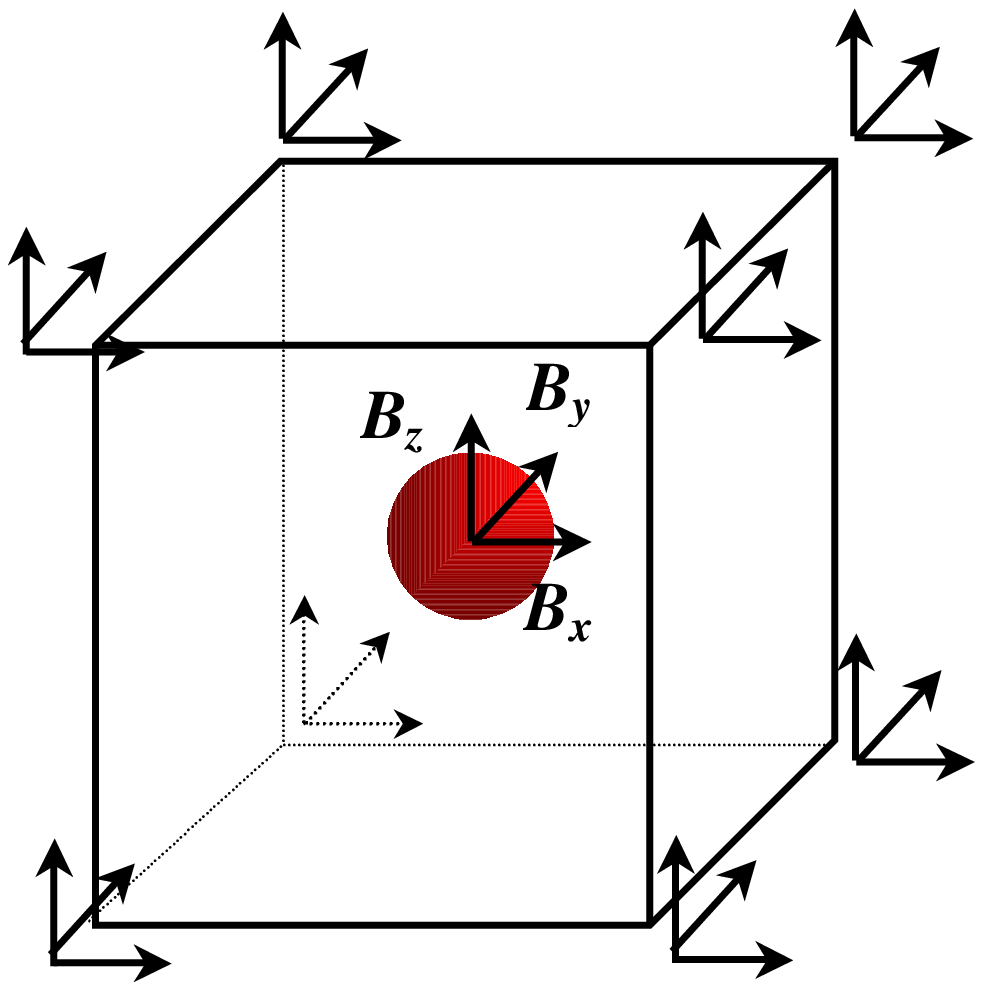,width=8cm,clip=}
\end{center}
\caption{Schematic representation of the probe geometry around the
  vacuum  cell. The compensating coils are not shown.}
\label{fig:Cell}
\end{figure}

In practice, we placed twenty-four magneto-resistive
probes at the eight corners 
of the vacuum cell, three at each corner oriented according to the three
orthogonal directions (see Fig. \ref{fig:Cell}).
The probes are mounted on two U-shaped integrated circuits placed
parallel to the top and bottom faces of the cell, preserving a large
optical access to the cell. A maximum of algebraic operation is
performed directly on the integrated circuits in order to reduce the
number of external electric connections.
The sum of the eight signals for a given direction generates three error
signals: Each error signal is
integrated and the result is used to drive a
current supply (we simply use standard diode laser supplies in our setup)
{\em via} a modulation input. Each supply drives
a coil pair in Helmholtz configuration generating a magnetic field
opposite to that measured by the probes.
The rectangular compensating coils are typically 1.5
meter wide in order to insure the homogeneity of the compensated
magnetic field across
the vacuum cell (whose size is 10 cm). The magneto-resistive probes
(HMC2003 Honeywell) have, according to the manufacturer data, a 
resolution better than 40 $\mu G$. In closed-loop operation, the
residual magnetic field
fluctuation measured by the probes is 150 $\mu G$,
corresponding to a reduction of one order of magnitude,
and the bandwidth is better than 500 Hz.
We took care of aligning the probes axes with the axes of the
compensation coils to better than five degrees, and we never saw any
instabilities due to the coupling among the three directions with
the three servo-loops acting simultaneously.
With this setup, canning the DC magnetic field component along one direction
(by scanning the respective DC current) does not affect
the stability of the compensation along the other directions,
even in presence of unavoidable stray couplings among directions,
which are correctly compensated by the device.

A good way of testing the effectiveness our setup is to perform
{\em co-propagating} Raman spectroscopy.
We have seen in Sec.~\ref{sec:Intro} that such a configuration
is almost insensitive to
the atomic velocity, and it allows easy diagnostic of perturbing effects
like residual stray magnetic fields and light-shifts.
To do so, we programmed our experiment to execute the following
time sequence: The magneto-optical (MOT) trap is 
loaded; the MOT gradient of magnetic field is shut down and a
``Sisyphus'' sequence is applied for further cooling.
Twenty millisecond after cutting the MOT gradient, the magneto-resistive
probes are reset. Reseting must be done before the probes can perform
measurements once they have been submitted to intense magnetic fields
(e. g. the MOT gradient).
A special function is provided in the probe chip for 
this end. A few millisecond later, the servo loops are switched on.
After twenty milliseconds, the optical beams are shut down, the atoms
are transfered form the $F=4$ to the $F=3$ sublevel and the
Raman pulse is applied. The Sisyphus sequence 
lasts for around 60 ms which is longer than in other cold atom
experiments (typically 30 ms), but is much shorter than the requested
delay (around 200 ms) when a mu-metal shield is used, necessary to
allow for the shield magnetization relaxation.

By applying a DC magnetic field during the experiment,
one can perform spectroscopy of cesium ground-level Zeeman states. 
If the DC field has no particular orientation with respect to the
polarizations of the Raman beams \cite{note:RamanPolarizations},
the selection rules for the Raman transitions
are $\Delta m=0,\pm 1,\pm 2$. As the Zeeman coefficient
for $F=3$ and $F=4$ are almost equal and have opposite signs,
$Z_3=350$ kHz/G and $Z_4=-351$ kHz/G, the position (with respect to
the line center) of the line connecting the sublevels
$(3,m)$ and $(4,m^\prime)$ is
\begin{equation}
\Delta \nu(m,m^\prime) \text{(kHz)} =
\left[ 350 (m+m^\prime)+m^\prime \right] B \text{(G) .}
\end{equation}

\begin{figure}
\begin{center}
 \epsfig{figure=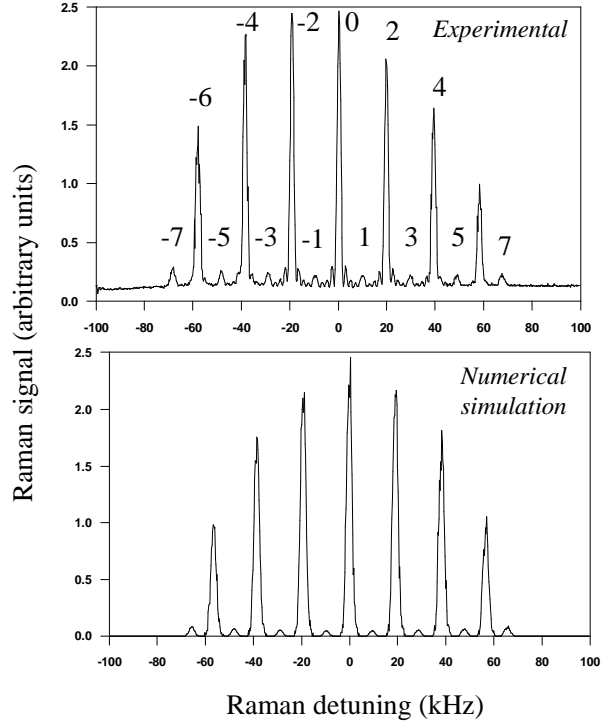,width=8cm,clip=}
\end{center}
\caption{Experimental and numerically generated
  (see Sec. \protect \ref{sec:Numerical}) Zeeman spectra of Raman
  transitions in cesium. The lines of small intensity
  correspond to transitions
$\Delta m = \pm 1$, showing that the DC field is almost parallel
to the Raman beams wave vector ${\bf k}$. The numerical simulation was
made with $B_0 = 28$ mG, $\Delta B_0 = 0.5$ mG, $\Delta I_0/I_0 = 0.025$
and $\theta_0 = 0.2$ rad. The asymmetry of the experimental spectrum
is due to the fact that the Zeeman sublevels of the initial level are
not equally populated.}
\label{fig:DCSpectrum}
\end{figure}

\begin{figure}
\begin{center}
  \epsfig{figure=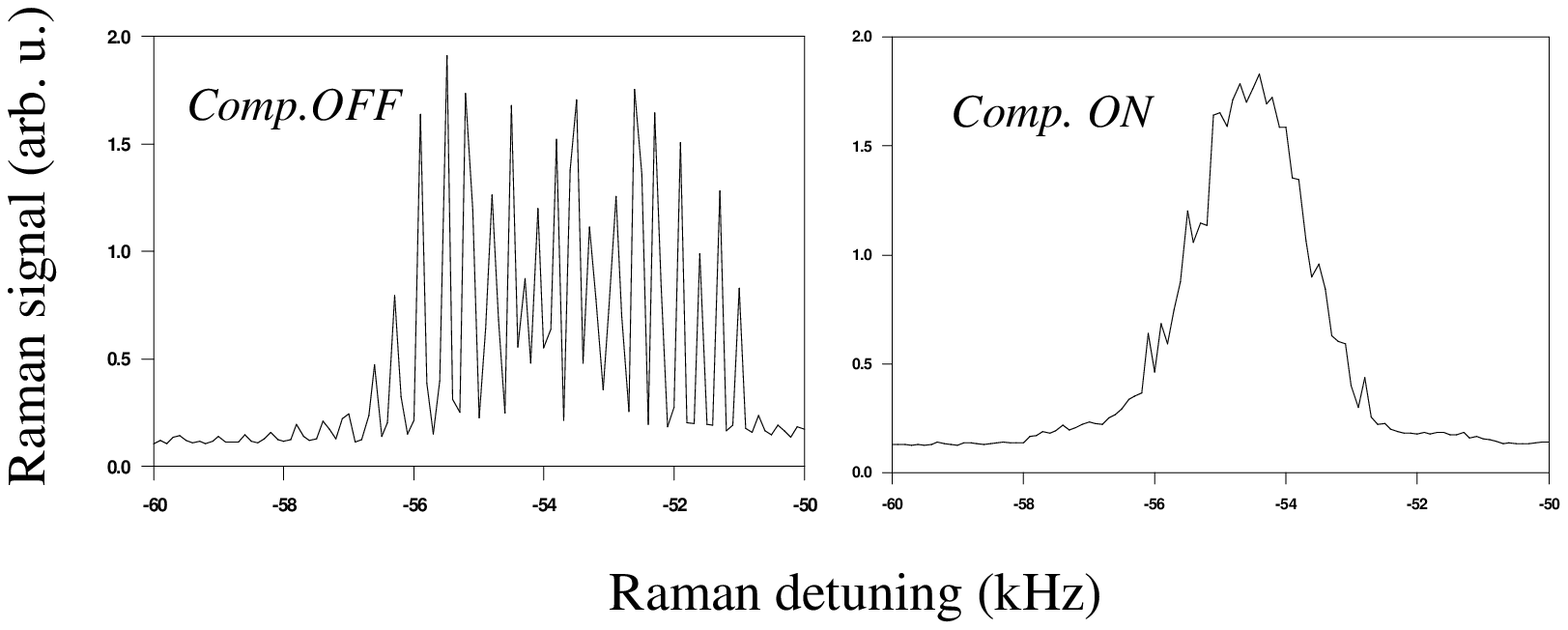,width=10cm,clip=}
\end{center}
\caption{ Detail of the line -6 of Fig. \protect
  \ref{fig:DCSpectrum} with the
  active compensation off and on. The non compensated curve presents a
  noise level close to 100\%.
  The regular oscillations superimposed to the line itself are
  due to a stroboscopic effect between the network frequency (50 Hz)
  and the repetition rate of the measurements (4 Hz).
  }
\label{fig:Zoom}
\end{figure}

As we are dealing with weak magnetic fields, the second term in the
brackets is in general negligible, and we can consider that the
position of the lines depends only on $m+m^\prime$. One easily deduces
that the spectrum can display up to 15 lines, as shown in Fig.
\ref{fig:DCSpectrum}. This figure has been
obtained by applying a DC field of 28 mG with the magnetic field compensation
setup described above active. The Raman pulse duration is 500 $\mu$s.
In Fig. \ref{fig:Zoom} we compare a ``zoom" of the line noted -6 in
Fig. \ref{fig:DCSpectrum} with the magnetic field active compensation on and
off, clearly showing a noise reduction effect. When the compensation
is off, only the DC component remains.

The maximum measured full width at half maximum (FWHM)
of the lines in Fig. \ref{fig:DCSpectrum} are around 1.5 kHz, which
means that the
resolution is independent of the magnetic sublevel and limited by the
Fourier transform of the Raman pulse.
For longer interaction times, we have observed a linewidth of 600 Hz
which correspond to a velocity resolution better than $v_r /10$ and
thus to a field residual amplitude of 300 $\mu G$.
To achieve this last value, we added a gradient-compensating
coil pair in anti-Helmholtz configuration, that generates a 10 mG/cm
gradient to eliminate residual MOT-gradient magnetic fields.

\begin{figure}
\begin{center}
  \epsfig{figure=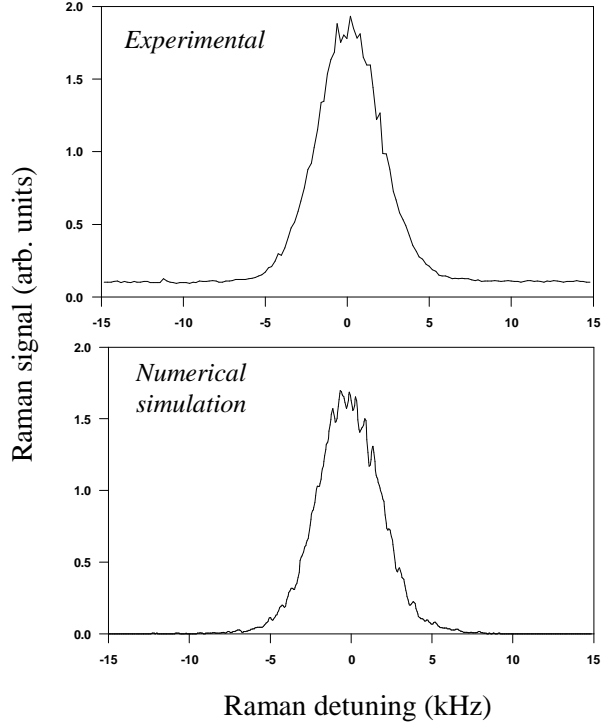,width=8cm,clip=}
\end{center}
\caption{Collapsed Raman spectrum at low magnetic field (FWHM 4.5 kHz).
Comparison with the corresponding spectrum obtained 
by numerical simulation (see Sec. \protect \ref{sec:Numerical}),
obtained with $B_0 = 0.5$ mG, $\Delta B_0 = 0.3$ mG, $\Delta I_0/I0 = 0.027$
and $\theta_0 = 0.01$ rad. }
\label{fig:CollapsedSpectrum}
\end{figure}

Reducing the DC field leads to a collapsed, Zeeman-compensated
spectrum, with a FWHM of about 4 kHz (see Fig.
\ref{fig:CollapsedSpectrum}). In principle, one would expect the width of the
collapsed spectrum to be just slightly greater than the widest line of
the split spectrum. Explaining this ``unexpected" broadening
is the purpose of the next section.

\begin{figure}
\begin{center}
  \epsfig{figure=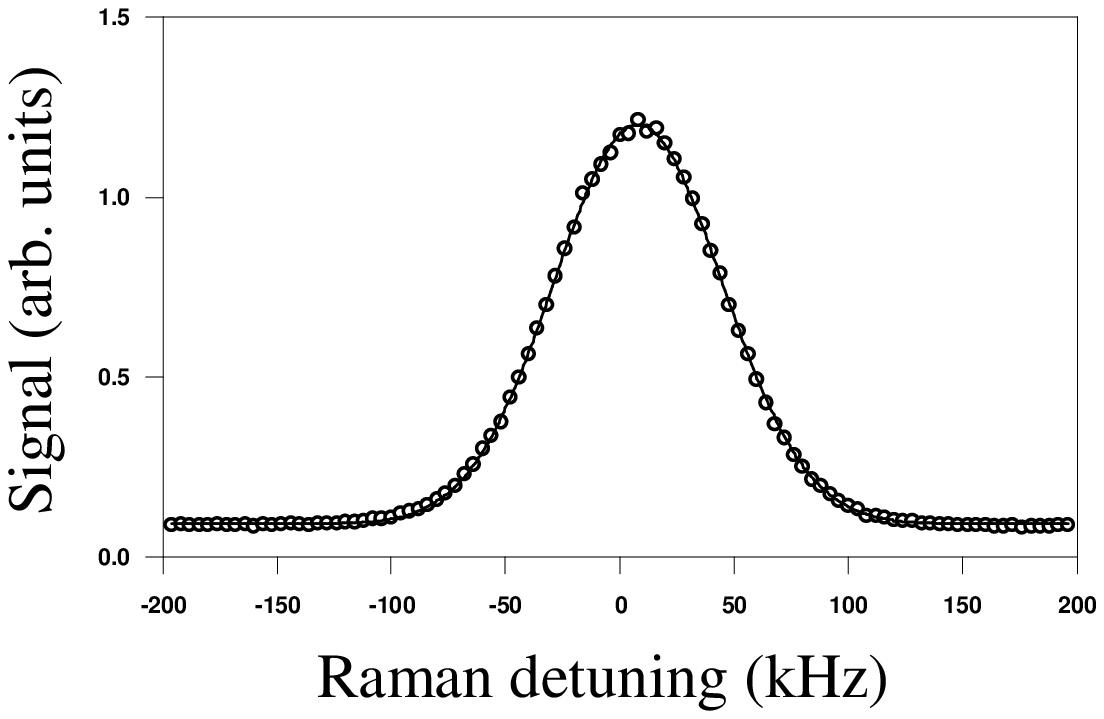,width=8cm,clip=}
\end{center}
\caption{Velocity distribution of cooled cesium atoms. The FWHM is 82 kHz,
corresponding to a temperature of 3.3 $\mu K$, deduced from the gaussian
fit. The resolution is of $v_r/2$ (or, equivalently, 4.2 kHz), showing
the ability of the setup to perform subrecoil spectroscopy.}
\label{fig:Contraprop}
\end{figure}

With the performances described above, we easily achieved a
velocity-measurement resolution of $v_r/2$. As an example,
Fig. \ref{fig:Contraprop} present a measurement of the velocity
distribution of the cesium atoms. The FWHM is 82 kHz, corresponding
to a temperature of 3.3 $\mu$K which is typical of ``Sisyphus-boosted"
MOTs. This proves the ability of the setup to perform subrecoil spectroscopy.

Concluding this section, we can say that this setup displays a
sufficient resolution to be used in most Raman stimulated spectroscopy
experiments. Furthermore, there is hardly any reduction of the optical
access to the vacuum cell.
Let us however mention that the setup presents a serious limitation in
the everyday use. We observed that the HMC2003 probes have a slow
time-drift, with a time constant of roughly one hour.
This implies that the DC component of the magnetic field
compensation has to be readjusted regularly.
We accounted for this effect by a software procedure 
restoring a fixed value of the magnetic field (as measured by the
probes), for instance each hour. We also 
tested the possibility of replacing the magneto-resistive probes by small 
coils and use the induction signal to generate an AC error signal.
This technique apparently presents no detectable drift.

\section{Study of the spectra}
\label{sec:Numerical}

The relative intensity of the Raman transitions depends on the Raman amplitude
connecting the respective Zeeman sublevels, which is proportional to
the product of the dipole matrix elements connecting each sublevel to
the concerned intermediate excited level (see Appendix \ref{app:formulas}).
The position of each Zeeman sublevel depends on the magnetic field
and on the intensity of the Raman beams through the light-shift, which,
for the far detuned Raman beams we are considering here, depends only on
the dipole matrix element connecting the considered Zeeman sublevel to the
excited level. 
We choose the magnetic field ${\bf B}$ to be parallel to the quantization
axis, labeled ${\bf z}$. The Raman beams are 
co-propagating with wavevector ${\bf k}$ (see Fig. \ref{fig:Geometry}), and
lie in the $(x-z)$ plane, making an angle $\theta$ with respect
to the ${\bf z}$-axis. The (orthogonal) polarization vectors
associated with the two Raman beams lie in a plane orthogonal to
${\bf k}$, and the polarization of the Raman beam starting from the level
$F=3$ makes an angle $\varphi$ with the $(x-z)$ plane (this last angle is
irrelevant in the weak magnetic field limit, we shall thus always
take $\varphi = 0$ in what follows).

We consider three main causes of line broadening: (a) Magnetic 
field fluctuations and inhomogeneities, (b) fluctuations and
inhomogeneities of the intensity of the Raman beams, and
(c) the finite duration of the
pulses. We shall consider each of these effects in detail.

The fluctuations of the magnetic field displace the Zeeman sublevels,
and thus displace the center of each Raman
line (except the line connecting the Zeeman $m=m^\prime=0$ sublevels, if
allowed). Furthermore, we consider that the magnetic field in our
setup results from the vectorial addition of a controlled, 
DC vector ${\bf B}_0$ and of a randomly fluctuating component
${\bf \Delta B}$. The resulting magnetic field  thus varies
both in modulus and in direction. The quantization axis is aligned
with the DC component. The spatial inhomogeneity of the
magnetic field also displaces the position of the Raman lines,
thus contributing to the
inhomogeneous broadening. We do not know the exact distribution of the
magnetic field intensity over the atomic cloud, but as the atoms have
an almost random spatial distribution, this contribution can be
statistically treated in the same way as a field
fluctuation and we do not distinguish between them.
The typical magnitude of this effect in the
weak magnetic field regime is 1.5 kHz.

The dominant low-magnetic field broadening factor 
is the fluctuation of the light-shifts (as for the magnetic field effect,
we do not distinguish the contributions of fluctuations and of
inhomogeneities). These fluctuations
are in part due to the fluctuations of the intensity of the Raman
beams themselves. However, the light-shift affecting a given Zeeman
sublevel $(F,m)$ is {\em independent of $F$
and $m$ if the direction of the magnetic field is parallel to the direction of
propagation of the Raman beams} (this result is also valid for zero
magnetic field, as one can then always chose the quantization axis
parallel to the beams). This implies that the light-shift fluctuations
in this configuration displace the lines as a whole,
and cannot contribute to the broadening of an individual line.
This fact shall play an essential role in the following of this
section.
Thus, we do not expect the best compression of
the spectrum to be achieved with zero DC field, because
in this case the fluctuating component
${\bf \Delta B}$  induces magnetic field
random rotations with respect to ${\bf k}$, leading to spectra that
are broad and asymmetric. The thinnest lines are expected to be
obtained with a DC field 
parallel to the Raman beams of an amplitude a few times greater than the mean
amplitude of the fluctuating component: in such case, the fluctuation
of the magnetic field induce only small misalignments with respect to
propagation direction of the Raman beams. We verified these facts both
experimentally and numerically (see below). Let us mention that a
simple way to obtain the alignment of the DC field and the Raman
beams is to try to minimize the amplitude of the
lines corresponding to $\Delta m=\pm 1$
(that is, -7,-5,..,5,7) in the spectrum of Fig. \ref{fig:DCSpectrum},
as the Raman selection rules forbid such transitions
if $\bf{B}$ and ${\bf k}$ are parallel. If the axes of the
compensation coils are aligned with the
Raman beams (as it is the case in our setup), one can reduce the DC
component without loosing the correct field orientation.

Lowering the intensity of the Raman beams produces thinner
linewidths, but the overall signal also decreases, specially
when performing counter-propagating (velocity-sensitive) Raman
spectroscopy. The intensity used in our quantum chaos experiment
\cite{ref:Bicolor} and in the curves shown in the present paper is
20 mW for each Raman beam of waist 4 mm. The typical value of the
broadening associated with the light-shift fluctuations is 3.5 kHz.

The last broadening effect is the finite duration of the Raman
pulses. We worked with Raman pulses ranging from 1 to 3 ms for which
the associated broadening (around 0.2 kHz) is negligible.

Numerically calculated spectra have been generated by a Monte-Carlo technique.
A particular value for the magnetic field is generated by adding a
DC component ${\bf B_0}$ making an angle $\theta_0$ with ${\bf k}$
and a perturbation${\bf \Delta B}$, whose amplitude is picked
from a gaussian distribution of zero mean, root-mean-square
amplitude $\Delta B_0$, and of random orientation. Note that $\Delta
B_0$ is formed by two contributions: the stray local fields
(independent of $B_0$) and the fluctuations of $B_0$ itself, which
we assume to be  proportional to $B_0$ (due e. g. to fluctuations
of the current delivered by the supplies).

In the same way, a value for the 
intensity of each Raman beam is picked from a gaussian distribution of
mean intensity $I_0$ and root-mean-square variation $\Delta I_0$.
The position and the amplitude of each
line is then calculated, and a particular spectrum generated using
the ``natural'' linewidth given by the inverse of the pulse duration.
Typically 1000 of such randomized spectra are averaged to produce
the spectra shown in
Figs. \ref{fig:DCSpectrum} and \ref{fig:CollapsedSpectrum}. 

\begin{figure}
\begin{center}
  \epsfig{figure=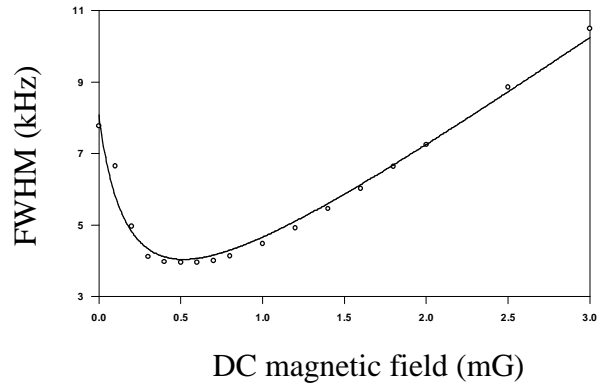,width=8cm,clip=}
\end{center}
\caption{Theoretical FWHM of the collapsed spectrum as a function of
  the applied DC field. The fitting curve is
  $1.74/(B_0+0.21)+3.24 B_0$, with $B_0$ in mG.}
\label{fig:FWHMxB0}
\end{figure}

As expected, the best collapsed spectrum is obtained for a
non-zero value of the DC field. Fig. \ref{fig:FWHMxB0} displays the
numerically calculated dependence of the FWHM on the DC magnetic field
$B_0$, showing a well defined minimum around $B_0 = 0.5$ mG. For
stronger mean fields, the broadening is dominated by the fluctuations
of $B_0$ itself, and should be linear ion $B_0$. For values $B_0$ close
to zero, the broadening is due essentially to field rotations induced by the
fluctuation component. The fluctuations of the field
direction are roughly proportional to $\Delta B_0/B_0$ and the related
broadening is maximum if $B_0 = 0$, but it cannot be infinite. We thus
write the weak field term as $A/(B_0+b)$, where $A/b$ is the zero DC
field broadening, depending only on $\Delta B_0$.
The shape of the curve is then of the form
\begin{equation}
  F(B_0) = { A \over B_0+b } + C B_0 \text{ .}
  \label{eq:fitting}
\end{equation}
Fig. \ref{fig:FWHMxB0} shows that this forms fits rather well the
numerical simulation. Unfortunately, we cannot directly compared this
result to experimental data as we do not have an independent measurement of
the mean magnetic field across the atomic cloud.

Interestingly, one also finds that the
minimum width does not correspond to $\theta=0$ (perfect alignment 
of the magnetic field and the Raman beam wavevector) but to a small angle
$\theta_0 = 0.1$ rad. For such an angle, light-shifts are
slightly different for each Zeeman sublevel, and so are the Zeeman
shifts. We attribute the minimum to a partial compensation of the
two effects. However, this misaligned configuration
tends to produce slightly asymmetric spectra.

Concluding this section, we can say that we have provided a
convincingly interpretation of the physical mechanisms governing the
width of the Raman lines. The interpretation has been tested by
comparing
numerically generated spectra to the experimental results, which
allowed us to quantify the performances of our setup.
Moreover, this interpretation furnishes some
interesting hints for the minimization of the width of the
collapsed spectra. Another interesting hint is to polarize the atoms
in one of the extreme ($|m|=F$) Zeeman sublevels by optical pumping.
This shall sensibly reduce the broadening due to fluctuations of the
light-shift. Based on the measurements we have done in the split spectrum
configuration, we expect in this case a collapsed linewidth of the
order of 1 kHz.

\section{Conclusion}
\label{sec:Conclusion}

We presented in this paper an experimental setup for subrecoil
stimulated Raman spectroscopy. A key feature of the setup is the active
compensation of magnetic fields over a small volume. Sensitive
Raman stimulated transitions on cesium have
been used to characterize the setup, with residual magnetic field
fluctuations of 
$0.3$ mG. The final
limit of the method has been shown to be related to light-shift
fluctuations rather than to the magnetic field ones.
The residual linewidth is 4 kHz, provided a small non-zero DC field is
kept, allowing atomic velocity selection below $v_R/2$ with the
present method. Carefully constructed
Raman $\pi$-pulses, would allow to reduce the light power, pushing the
resolution on the atomic velocity selection
to the $v_R/10$ level, which would be comparable to the results
obtained with double mu-metal shielding 
\cite{ref:BlochOsc}.

Laboratoire de Physique des Lasers, Atomes et
Mol{\'e}cules (PhLAM)
is UMR 8523 du CNRS et de l'Universit{\'e} des Sciences
et Technologies de Lille. Centre d'Etudes et Recherches
Lasers et Applications (CERLA) is supported by Minist\`{e}re de la
Recherche, R\'{e}gion Nord-Pas de Calais and Fonds Europ\'{e}en de
D\'{e}veloppement Economique des R\'{e}gions (FEDER).

\appendix
\section{Light-shifts and transition rate coefficients}
\label{app:formulas}
\begin{figure}
\begin{center}
  \epsfig{figure=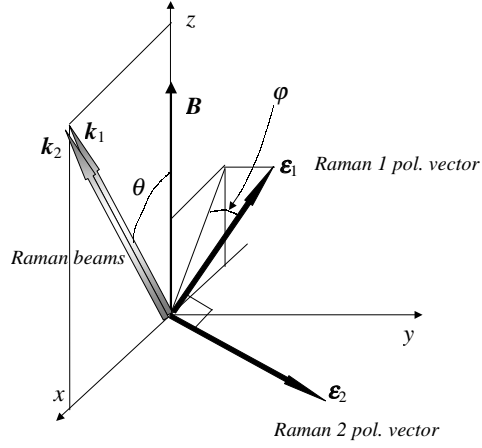,width=8cm,clip=}
\end{center}
\caption{Geometry of the Raman beams and the magnetic field.}
\label{fig:Geometry}
\end{figure}

The geometry and definitions used in this appendix are those of Fig.
\ref{fig:Geometry}, explained in the preceding section. 
We choose the quantization axis parallel
to the magnetic field and decompose the polarizations of the Raman
beams in irreducible components (noted
$\bf{\epsilon}_-,\bf{\epsilon}_0,\bf{\epsilon}_+$) 
with respect to that axis \cite{note:RamanPolarizations} 

\begin{eqnarray}
  \epsilon_{1-} = { -\cos \varphi \cos \theta + i \sin \varphi \over
    \sqrt{2} } \nonumber \\
\epsilon_{10} = \cos \varphi \sin \theta \nonumber \\
\epsilon_{1-} = { \cos \varphi \cos \theta + i \sin \varphi \over \sqrt{2} }
\label{eq:polarization}
\end{eqnarray}
and the components of the polarization vector $\bf{\epsilon}_{2}$
of the second beam are
obtained by changing $\varphi$ into $\pi/2-\varphi$.

The light-shift of the sublevel $(F,m)$ induced by the beam labeled
$i$ ($i=1,2$) is proportional to
\begin{equation}
L_i(F,M) = (qE_i)^2 \sum_{F^\prime} 
{ | \sum_p \langle 6S_{1/2} F m|r_p \epsilon_{ip}|6P_{3/2} F^\prime
  m-p  \rangle |^2
\over \Delta_i(F,M;F^\prime,M^\prime) }
\label{eq:Light-shift}
\end{equation}
where $q$ is the electron charge, $E_i$ the electric field associated
to the $i$ Raman beam, $p=\{-,0,+\}$, $r_p$ is the irreducible
$p$-component of the position vector, and $(F^\prime,m^\prime)$
correspond to the Zeeman sublevels of the excited state
($F^\prime = 2,3,4$ if $F=3$ and $F^\prime = 3,4,5$ if $F=4$), and 
$\Delta_i(F,m,F^\prime,m^\prime)$ is the optical detuning. For
simplicity, we will consider that the optical detuning
is high enough so that the
above value can be considered to be independent of the sublevel (practically, 
this means that the optical detuning must be much larger than 250
MHz). Applying the Wigner-Eckart theorem, one can reduce the above
equation to the more useful form:
\begin{eqnarray}
  L_i(F,m)= (qE_i)^2 \sqrt{2} (2F+1) (2J+1) { |
      \langle 6S_{1/2}||{\bf r}||6P_J \rangle |^2
    \over \Delta } \times \nonumber \\
\sum_{F^\prime} \left| \sum_p \epsilon_{ip} \sqrt{2F^\prime+1}
\left\{
\begin{array}{ccc}
   1/2      & J & 1 \\
   F^\prime & F & 7/2
\end{array}
\right\}
\left(
\begin{array}{ccc}
   F^\prime & 1 & F \\
   m-p      & p & -m
\end{array}
\right)
\right|^2
\end{eqnarray}
where the symbol into brackets is the reduced matrix element of $\bf{r}$, 
the symbol into braces a ``6j" symbol and the symbol into parenthesis a ``3j"
\cite{ref:Landau}. The total light-shift $L(F,m)$ affecting the sublevel 
$(F,m)$ is thus obtained by adding the contributions due to each Raman beam.

Let us now consider the Raman transition amplitude $(F_1,m_1) \rightarrow
(F_2,m_2)$, which is proportional to
\begin{equation}
\sum_{F^\prime} \sum_p \sum_{p^\prime}
 \langle 6S_{1/2} F_1 m_1|r_p \epsilon_{1p} |6P_{3/2} F^\prime m_1-p \rangle
 \langle 6P_{3/2} F^\prime m_1-p | r_{p^\prime} \epsilon_{2p^\prime}
 | 6S_{1/2} F_2 m_1-p-p^\prime \rangle
\label{eq:RamanTransAmpl}
\end{equation}
that can be transformed into:
\begin{eqnarray}
  \nonumber
  |\langle 6S_{1/2} || {\bf r} || 6P_J \rangle|^2  (-1)^{F_1+F_2-q-J+1/2}
  \sqrt{2 (2F_1+1) (2F_2+1) (2J+1)}\\
  \nonumber
  \sum_p \sum_p^\prime \epsilon_{1p} \epsilon_{2p^\prime}
  \left \{  \begin{array}{ccc}
    0 & 1 & 1 \\
    J & 1/2 & 1/2
  \end{array} \right \}^2
\sum_{F^\prime} (2F^\prime+1)
\left \{ \begin{array}{ccc}
    1/2 & J & 1 \\
    F^\prime & F_1 & 7/2
  \end{array} \right \}
\left \{ \begin{array}{ccc}
    J & 1/2 & 1 \\
    F_2 & F^\prime & 7/2
  \end{array} \right \} \\
\left( \begin{array}{ccc}
    F^\prime & 1 & F_1 \\
    m_1-p & p & -m_1
  \end{array} \right)
\left( \begin{array}{ccc}
    F_2 & 1 & F^\prime \\
    m_1-p-p^\prime & p^\prime & -(m_1-p)
  \end{array} \right)
\end{eqnarray}

\vspace{1cm}

\vspace{0.5cm}

\vspace{0.5cm}

\vspace{0.5cm}

\vspace{0.5cm}

\vspace{0.5cm}

\vspace{0.5cm}

\vspace{0.5cm}

\vspace{0.5cm}

\vspace{0.5cm}

\begin{references}

\bibitem{ref:Subrecoil} M. Kasevich and S. Chu, Phys. Rev. Lett.
  {\bf 69}, 1741 (1992).
\bibitem{ref:BlochOsc} M. Ben Dahan, M. Pike, J. Reichel, Y. Castin,
  and C. Salomon, Phys. Rev. Lett. {\bf 76}, 4508 (1996).
\bibitem{ref:Sideband} V. Vuletic, C. Chin, A.J. Kerman, and S. Chu,
Phys. Rev. Lett. {\bf 81}, 5768 (1998);
M. Morinaga, I. Bouchoule, J. C. Karam, and C. Salomon, 
Phys. Rev. Lett. {\bf 83}, 4037 (1999).
\bibitem{ref:Bicolor} J. Ringot, P. Szriftgiser, J. C. Garreau, and D. Delande,
Phys. Rev. Lett. {\bf 85}, 2741 (2000).
\bibitem{ref:DiodeModulee} J. Ringot, Y. Lecoq, J. C. Garreau,
  and P. Szriftgiser, Eur. Phys. J. D {\bf 7}, 285 (1999).
\bibitem{note:geometry}
  Nevertheless, this is valid only for the exact geometric center of
  the cube. If the magnetic gradient is too high,
  the spatial extension of the atomic cloud can lead
  to an important broadening due to the inhomogeneity of the residual
  magnetic field.
\bibitem{note:RamanPolarizations} We work with Raman beams of
  orthogonal  polarizations, as, for beams far detuned from the
  optical resonance, the transition probability with parallel
  polarizations is zero. This is due to symmetry properties of
  the Clebsch-Gordan coefficients when summed of all hyperfine
  sublevels of the intermediate level $6P_{3/2}$.
\bibitem{ref:Landau} Definitions and very useful formulas for calculation of
these symbols can be found in the chapter 14 of Landau and Lifshitz,
\it{ M{\'e}canique Quantique, Th{\'e}orie non relativiste}, Mir,
Moscou, 1966.
Moreover, formulas (106,14), (108,10) and (106,10) are well adapted for the
numerical computation of ``3j'', ``6j'' symbols and Clebsch-Gordon
coefficients.
\end{references}
\end{document}